# Comment on
# "Radiation effects on a certain MHD free convection flow", by Ahmed Y. Ghaly [Chaos, Solitons & Fractals, 13 (2002) 1843-1850]

Asterios Pantokratoras
Associate Professor of Fluid Mechanics

School of Engineering, Democritus University of Thrace,
67100 Xanthi – Greece
e-mail:apantokr@civil.duth.gr

In the above paper the author treats the boundary layer flow along a vertical isothermal plate of an electrically conducting fluid. The plate lies in a vertical free stream of constant velocity $u_\infty$, constant temperature $T_\infty$ and constant mass concentration $C_\infty$. The plate moves vertically with a velocity which is a linear function of distance x ($u_w$=cx). A transverse magnetic field with constant strength $B_0$ is imposed on the flow. The boundary layer equations are integrated numerically. However, there is a fundamental error in this paper and the presented results do not have any practical value. This argument is explained below:

The momentum equation used by the author is

$$u\frac{\partial u}{\partial x} + v\frac{\partial u}{\partial y} = \nu\frac{\partial^2 u}{\partial y^2} - \frac{\sigma B_0^2}{\rho}u + g\beta(T - T_\infty) + g\beta^*(C - C_\infty) \qquad (1)$$

where u and v are the velocity components, ν is the fluid kinematic viscosity, ρ is the fluid density, σ is the fluid electric conductivity, g is the gravity acceleration, T is the fluid temperature, C is the mass concentration, β is the fluid thermal expansion coefficient and $\beta^*$ is the fluid concentration expansion coefficient.

The boundary conditions are:

at y = 0:   $u_w$ = cx,  v = 0, T = $T_w$, C=$C_w$ \qquad (2)

as y → ∞   u = $u_\infty$, T = $T_\infty$, C=$C_\infty$ \qquad (3)

Let us apply the momentum equation at large y. At large distances from the plate the fluid temperature is equal to ambient temperature and the buoyancy term gβ(T-T$_\infty$) in momentum equation is zero. At large distances from the plate the mass concentration is also equal to ambient concentration and the buoyancy term gβ$^*$(C-C$_\infty$) is also zero. Taking into account that, at large distances from the plate, velocity is everywhere constant and equal to u$_\infty$ the velocity gradient $\partial u/\partial y$ is also zero. The same happens with the diffusion term $\nu \partial^2 u/\partial y^2$. This means that the momentum equation takes the following form at large y

$$u_\infty \frac{\partial u_\infty}{\partial x} = -\frac{\sigma B_0^2}{\rho} u_\infty \qquad (4)$$

or

$$\frac{\partial u_\infty}{\partial x} = -\frac{\sigma B_0^2}{\rho} \qquad (5)$$

Taking into account that the electric conductivity, density and the strength of the magnetic field are nonzero quantities the free stream velocity gradient is nonzero and the free stream velocity should change along x. However this finding is in contradiction with the assumption made in the paper that the free stream velocity is constant. Taking into account the above argument it is clear that the results presented by Ghaly (2002) are inaccurate.

REFERENCES

1. Ahmed Y. Gahly (2002). Radiation effects on a certain MHD free-convection flow, Chaos, Solitons & Fractals, Vol. 13, pp. 1843-1850.

Asterios Pantokratoras
Associate Professor of Fluid Mechanics
School of Engineering, Democritus University of Thrace,
67100 Xanthi – Greece
e-mail:apantokr@civil.duth.gr

In the above paper an analysis has been carried out to obtain results in a steady laminar boundary layer flow over a motionless horizontal plate placed in a horizontal free stream. The plate temperature is constant and different from the free stream temperature. In addition it is assumed that mass transfer takes place along the plate accompanied by a chemical reaction of this substance. The fluid viscosity is a function of temperature while the other fluid properties are assumed to be constant. The boundary layer equations are transformed into ordinary ones and subsequently are solved using the Chebyshev finite difference method. However, there are some errors in this paper which are presented below:

1. In nomenclature the symbol $U_0$ is denoted as "velocity of plate" but in the present problem the plate is motionless.
2. The definition of the transverse similarity variable $\eta(x,y)$ in equation (10) is wrong.
3. The relation between the transverse velocity v and the stream function $\psi$ is wrong.
4. The definition of the Nusselt number on page 65 is wrong.
5. In page 64 it is mentioned that "the present work deals with application of a radically new approach to the computation of the boundary-layer equations in MHD flows". However the above work has no relation to Magnetohydrodynamics (MHD).
6. In the transformed energy (temperature) equation (13) the symbol n (different from η) appears and this symbol has been used in all

the tables and figures taking the value n=0.01. This quantity is denoted in the nomenclature as "rate of chemical reaction" but has not defined in the paper. The reader can not understand and interprete the results without knowing this quantity. If we suppose that this quantity really exists and is relevant to chemical reaction then it must be included in the mass transfer equation (14) where the chemical reaction takes place. In the temperature equation (13) no chemical reaction takes place and the inclusion of n in this equation is unreasonable.

7. In the transformed energy (temperature) equation (13) the Prandtl number appears in two terms and has been assumed constant across the boundary layer. All the presented results concern Pr=1. However, the Prandtl number is a function of viscosity and viscosity has been assumed a function of temperature whereas the other fluid properties are considered constant and independent of temperature. Taking into account that temperature varies across the boundary layer, the Prandtl number varies, too. The assumption of constant Prandtl number inside the boundary layer, with temperature dependent viscosity, is a wrong assumption and leads to unrealistic results (Pantokratoras, 2004, 2005, 2007). In these three paper by Pantokratoras the difference in the results between variable Prandtl number (correct assumption) and constant Prandtl number (wrong assumption) reached 435 % , 85 %.and 98 %. The problem can be treated properly either considering the Prandtl number as a variable in the transformed equations (Saikrishnan and Roy, 2003) or with the direct solution of the initial boundary layer equations and treating the viscosity as a function of temperature (Pantokratoras, 2004, 2005, 2007).

Taking into account the above arguments there are doubts for the credibility of the above work.